\documentclass[aps,amsfonts,amsmath,amssymb,nofootinbib,twocolumn,superscriptaddress]{revtex4}

\usepackage{graphics}
\usepackage{hyperref}
\usepackage{epsfig}
\usepackage{color}

\usepackage[normalem]{ulem}
\usepackage{multirow}
\usepackage{bm}

\newcommand{\red}[1]{{\textcolor{black}{#1}}}

\newcommand{\bvec}[1]{{\bm{#1}}}

\begin{document}

\bibliographystyle{naturemag}

\title{Magnetic field control of continuous N\'eel vector rotation and N\'eel temperature in a \red{van der Waals} antiferromagnet} 

\author{Zhuoliang Ni}
\affiliation{Department of Physics and Astronomy, University of Pennsylvania, Philadelphia, PA 19104, U.S.A}
\author{Urban Seifert}
\affiliation{Kavli Institute for Theoretical Physics, University of California, Santa Barbara
Santa Barbara, CA 93106, U.S.A.}
\author{Amanda V. Haglund}
\affiliation{Department of Materials Science and Engineering, University of Tennessee, Knoxville, TN 37996, U.S.A.}
\author{Nan Huang}
\affiliation{Department of Materials Science and Engineering, University of Tennessee, Knoxville, TN 37996, U.S.A.}
\author{David G. Mandrus}
\affiliation{Department of Materials Science and Engineering, University of Tennessee, Knoxville, TN 37996, U.S.A.}
\affiliation{Materials Science and Technology Division, Oak Ridge National Laboratory, Oak Ridge, TN, 37831, U.S.A.}
\author{Leon Balents}
\affiliation{Kavli Institute for Theoretical Physics, University of California, Santa Barbara
Santa Barbara, CA 93106, U.S.A.}
\author{Liang Wu}
\email{liangwu@sas.upenn.edu}
\affiliation{Department of Physics and Astronomy, University of Pennsylvania, Philadelphia, PA 19104, U.S.A}

\date{\today}

\begin{abstract}

In a collinear antiferromagnet, spins tend to cant towards the direction of an applied magnetic field, thereby decreasing the energy of the system. The canting angle becomes negligible when the magnetic field is small so that the induced anisotropic energy is substantially lower than the exchange energy. However, this tiny anisotropy can play a significant role when the intrinsic anisotropy of the antiferromagnet is small. In our work, we conduct direct imaging of the Néel vector in a two-dimensional easy-plane antiferromagnet, MnPSe$_3$, with negligible spin canting under an external in-plane magnetic field. The small inherent in-plane anisotropy allows for the continuous rotation of the Néel vector  by ramping up the magnetic field in samples from the bulk to the monolayer. In monolayer samples, the applied magnetic field elevates the Néel temperature 10$\%$ at 5 tesla, as the combination of intrinsic and field-induced anisotropies set a critical temperature scale for fluctuations of the otherwise disordered Néel vector field. Our study illuminates the contribution of field-induced anisotropy in two dimensional magnets with in-plane anisotropy. We also demonstrate that the strain can tune the spin flop transition field strength by one order of magnitude.

\end{abstract}

\maketitle


Antiferromagnetic (AFM) materials are often resistant to external magnetic fields because the magnon gap is often in the order of 10 Tesla \cite{baltzrmp2018}. When a magnetic field is applied perpendicular to the Néel vector in a collinear antiferromagnet, all spins tend to cant towards the magnetic field direction, lowering the steady-state energy. This is supported by the observation that the magnetic susceptibility has a finite value under a perpendicular magnetic field. In contrast, the magnetic susceptibility under a parallel magnetic field is zero ($T=0$), indicating no steady-state energy change. However, as one gradually increases the magnetic field parallel to the Néel vector, the Néel vector will  interestingly rotate to be perpendicular to the magnetic field when the energy difference between parallel and perpendicular situations surpasses the anisotropy energy. This is known as the spin-flop transition. Typically, the spin-flop transition field needs to be large enough to overcome the anisotropy energy in three-dimensional materials, ranging 1-42 T.\cite{Palacioprb1980,blazeypr1968,jaccarinojmmm1983,jacobsjap1961,Jacobspr1967}.

The discovery of two-dimensional antiferromagnets\cite{leenanoletter16,ninatnano21,Xzhangnanolett21,zhangnanolett21} offers new opportunities to investigate the interaction between magnetic fields and antiferromagnetically coupled spins\cite{songsci18,kleinsci18,wangnatmater21,longnanoletter20,basnetprm2021,feringaprb2022}, particularly in materials with weak anisotropy and strong fluctuation. Previous studies have shown a low spin-flop transition field in layered antiferromagnet CrCl$_3$ with in-plane spins\cite{mcguireprm2017,cainanoletter19,macneillprl2019,wangnatnano2019}. However, this transition has not been well-studied due to the lack of a direct probe for antiferromagnetic orders\cite{cainanoletter19,amilcarscience2021,wangnatnano2019,kleinnatphys19}. Additionally, the low exchange energy in layered antiferromagnets leads to significant spin canting, thus a net magnetization, resulting in a large deviation from antiferromagnetic behaviors\cite{wangnatnano2019}. In this work, we present our study on the antiferromagnetic states under magnetic fields in the van der Waals \red{easy-plane} magnet MnPSe$_3$ with very large exchange energy\cite{wiedenmannssc81,jeevanandamjpcm1999},. We will demonstrate how the in-plane magnetic field acts as an additional magnetic anisotropy and controls the magnetism of samples down to a monolayer by directly monitoring the continuous rotation of N\'eel vectors under the magnetic field, and over 10$\%$ increase \red{in the} Néel temperature, and the tunable spin flop field by one order of magnitude by the strain.

MnPSe$_3$ belongs to the AFM transition metal phosphorous trichalcogenide MPX$_3$ family (M=Mn, Ni, Fe, Co, V; X=S, Se)\cite{wiedenmannssc81,jeevanandamjpcm1999,wildesprb2006,ressoucheprb10,lanconprb16,Maisciadv2021,calderprb2021,YanAM2023}. Above the Néel temperature, MnPSe$_3$ belongs to the point group -3. Below the Néel temperature, a long-range ordering of Néel-type in-plane antiferromagnetism forms\cite{wiedenmannssc81}. Our previous studies have found that the orientation of the Néel vectors is easily controlled by the strain from the substrate\cite{ninatnano21}. No evidence of intrinsic (strain-free) easy-axis anisotropy has been reported, suggesting that \red{at low temperatures MnPSe$_3$ can be} described as an effective \red{easy-plane} magnet. \red{While widely varying estimates for the out-of-plane anisotropy have been reported in the literature \cite{wiedenmannssc81,jeevanandamjpcm1999,calderprb2021}, most recent studies suggest that the anisotropy is small compared to intrinsic exchange interactions \cite{calderprb2021,basnet22,jana23}.}

In the exchange limit, the energy density of a bipartite antiferromagnet \red{ with in-plane moments} like MnPSe$_3$ can \red{effectively} be expressed as \cite{baltzrmp2018}:
\begin{equation}
E=-2J_{ex}S^2 + 2K_u S^2 \sin^2(\beta - \theta),
\end{equation}
where $J_{ex}$ is the exchange energy, $K_u$ is the anisotropy energy, and $\theta$ and $\beta$ represent the orientation of the Néel vector and the direction of minimum anisotropy energy, respectively.
In the \red{easy-plane} magnet MnPSe$_3$, $K_u$ and $\beta$ are typically determined by external random strain field with arbitrary N\'eel vector direction and magnitude, and $K_u$ is much smaller than the exchange energy $J_{ex}$. To minimize the energy density, the Néel vector orientation $\theta$ must equal the direction of a minimum anisotropy energy. In situations where $K_u$ and $\mu_{0}H$ are much smaller than the exchange energy $J_{ex}$ (exchange limit), the magnetic field $H$ only induces negligible spin canting, proportional to $\frac{\mu_0\mu_BH}{J_{ex}}$. The primary contribution of the magnetic field is the induction of additional anisotropy energy. As shown in the supplementary information (SI), the energy density of the system under an in-plane magnetic field at zero degrees can be approximated as:
\begin{equation} \label{eq_KH}
E=-2J_{ex}S^2 + 2K_u S^2 \sin^2(\beta - \theta) - 2K_H S^2\sin^2(\theta),
\end{equation}
where $K_H=\frac{g\mu_0^2\mu_B^2H^2}{8J_{ex}S^2}$. The last term represents the additional anisotropy from the magnetic field. The anisotropy energy's magnitude is proportional to $H^2$, and it rotates the Néel vector towards the direction perpendicular to the magnetic field. Consequently, by applying an in-plane magnetic field, one can effectively tune the direction of the minimum magnetic anisotropy in the samples. \\

\textbf{Magnetic field controlled continuous Néel vector rotation}

\begin{figure*}
\centering
\includegraphics[width=\textwidth]{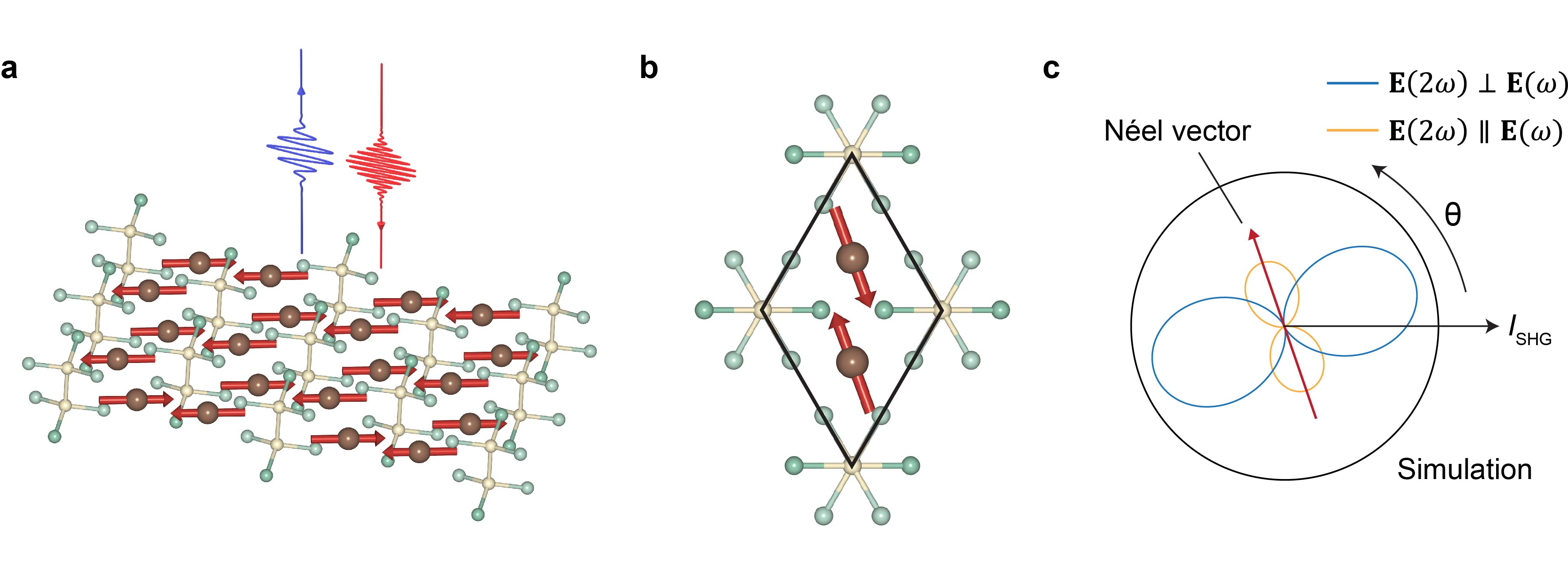}
\caption{\textbf{Optical Detection of Néel Vectors in MnPSe$_3$ using Second-Harmonic Generation.} {\bf a}, Depiction of the lattice and spin structure for a single MnPSe$_3$ layer. {\bf b}, Illustration of one MnPSe$_3$ monolayer unit cell. {\bf c}, Simulated polarization-resolved SHG corresponding to the Néel vector orientation shown in (b).}
\label{Fig1}
\end{figure*}

We employ second-harmonic generation (SHG) to monitor the Néel vector of MnPSe$_3$ under an in-plane magnetic field.
The Néel-type antiferromagnetic order breaks the inversion symmetry of the system, resulting in finite SHG signals\cite{fiebigprl94,ninatnano21,niprl2021, chuprl20,YanAM2023}. The magnitude and orientation of the Néel vector can be extracted using polarization-dependent SHG measurements\cite{ninatnano21}. Fig.\ref{Fig1}a illustrates the spin structure of a single MnPSe$_3$ layer with an arbitrary Néel vector orientation. A fundamental laser pulse is incident normally on MnPSe$_3$, and the reflected second-harmonic pulse is collected and measured. The unit cell of the Néel-type antiferromagnetism is depicted in Fig.\ref{Fig1}b. As previously reported the in-plane spin orientation of MnPSe$_3$ is determined by anisotropy from the strain\cite{ninatnano21}. Based on the spin structure in Fig.\ref{Fig1}b, the calculated polar patterns for the polarization dependence of SHG intensity from electric dipole contribution in parallel (${\bf E}(2\omega)\parallel {\bf E}(\omega)$) and crossed configurations (${\bf E}(2\omega)\perp {\bf E}(\omega)$) are displayed in Fig.\ref{Fig1}c \cite{ninatnano21}. \red{Note the relationship between the SHG polar patterns and the N\'eel vector has been proven in the previous study\cite{ninatnano21}}. The peak of the parallel configuration and the node of the crossed configuration indicate the Néel vector orientation, with their magnitudes being proportional to the square of the Néel vector magnitude\cite{ninatnano21}. If the Néel vector rotates, the SHG polar patterns will rotate by the same angle without altering the shape or magnitude\cite{ninatnano21}, \red{and therefore the SHG polar patterns are independent on the crystalline axis.} For a fundamental laser wavelength at 800 nm, the crossed pattern is an order of magnitude larger than the parallel pattern\cite{ninatnano21}; thus, we will only use crossed patterns to study the Néel vectors.

\begin{figure*}
\centering
\includegraphics[width=\textwidth]{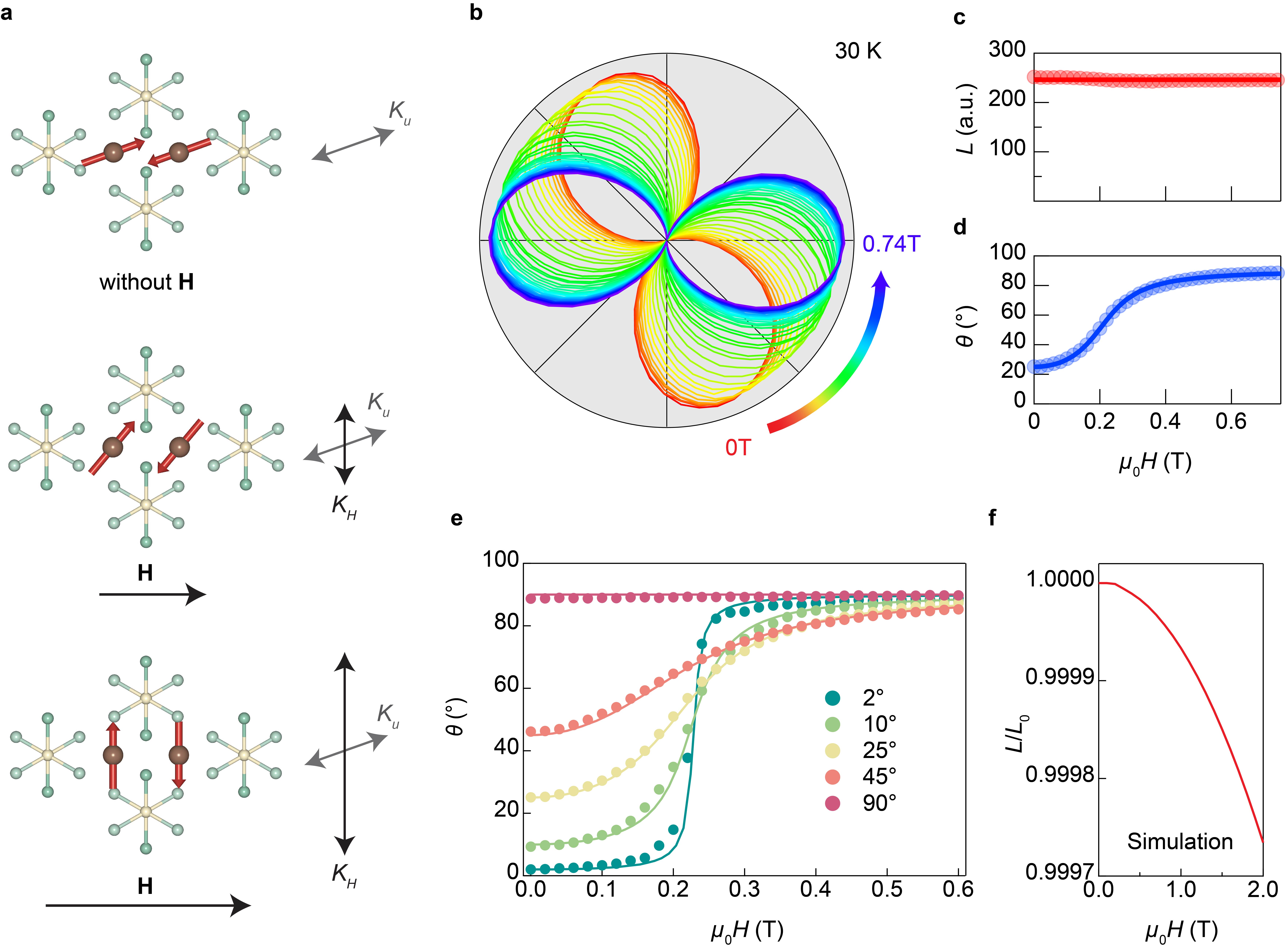}
\caption{\textbf{Magnetic Field Tuning of the Néel Vector in MnPSe$_3$.} {\bf a}, Schematic representation of the field-induced effective anisotropy energy $K_H$ rotating the Néel vector orientation. {\bf b}, Polarization-dependent SHG intensity measurements ($\mathbf{E}(2\omega)\perp\mathbf{E}(\omega)$) in a MnPSe$_3$ thin flake under varying magnetic fields. {\bf c}, Magnetic field dependence of the Néel vector magnitude, with circles representing experimental data and the solid line denoting simulation results. {\bf d}, Magnetic field dependence of the Néel vector orientation, with circles representing experimental data and the solid line denoting simulation results. {\bf e}, Néel vector orientation dependence on different magnetic field directions, with solid curves illustrating simulation results and dots representing experimental results. {\bf f}, Simulation of the Néel vector magnitude's response to the magnetic fields.}
\label{Fig2}
\end{figure*}

According to Eq.~\eqref{eq_KH}, an increasing in-plane magnetic field induces an increasing magnetic anisotropy energy, gradually rotating the Néel vector towards the direction perpendicular to the magnetic field. This process is illustrated in Fig.\ref{Fig2}a. We measure the polarization-dependent SHG pattern in a thin-flake MnPSe$_3$ sample of around 80 nm from 0 T to 0.74 T. The SHG polar patterns (in the crossed configuration) corresponding to the Néel vectors are presented in Fig.\ref{Fig2}b. We observe the node directions of the polar patterns gradually rotating from 25 degrees to 90 degrees, while the magnitudes remain constant. Here 0 degree is defined by the magnetic field direction.     By fitting the data, we extract the magnitudes and orientations of the Néel vectors as functions of the magnetic field and plot them in Fig.\ref{Fig2}c and d. The circles represent the experimental results, and the solid curves are derived by solving Eq.~\eqref{eq_KH}. The experimental data and theoretical results match well. 

In a different experiment, we maintain the magnetic field at zero degree, rotate the sample, and measure the Néel vector orientations as a function of the magnetic field. The experimental and theoretical results are depicted in Fig.\ref{Fig2}e. When the magnetic field is parallel to the original anisotropy direction, theoretically, no change in the Néel vector orientation occurs until the magnetic field reaches the spin flop transition and rotates the Néel vector by 90 degrees. In the \red{easy-plane} MnPSe$_3$, however, we observe a less pronounced change \red{green curve}. One possibility is that there is a small angle misalignment of 2 degrees between the  Néel vector and the magnetic field as shown in Fig.\ref{Fig2}e. Another possibility is that the small anisotropy from the external strain is not homogeneous. Consequently, the spin directions in each Mn atom under the laser beam spot exhibit a finite variance, resulting in a less abrupt transition when the magnetic field is nearly parallel to the average Néel vector directions. As we increase the angle between the magnetic field and the original anisotropy direction, the sharpness of the Néel vector orientation flop decreases. When the magnetic field is perpendicular to the original anisotropy direction, it has no effect on the Néel vector, as the additional anisotropy from the magnetic field shares the same direction as the original anisotropy.  To illustrate the negligible spin canting and verify our assumptions, we theoretically calculate the Néel vector magnitude as a function of the magnetic field, as shown in Fig.\ref{Fig2}f. At 1 T, the change in the Néel vector is below $10^{-4}$. Applying a magnetic field to an antiferromagnet usually causes spin canting, and, therefore, a drop of N\'eel vector magnitude occurs. Fig.\ref{Fig2}f represents the N\'eel vector magnitude as a function of the magnetic field. The small change indicates the spin canting is quite weak, which means at the spin-flop field MnPSe$_3$ is still antiferromagnet-like. This behavior makes MnPSe$_3$ unique because most antiferromangets have magnetic anisotropy close to the exchange energy, making the spin canting large at the spin-flop field.\\

\textbf{Strain tunable spin flop transition field}

\begin{figure*}
\centering
\includegraphics[width=\textwidth]{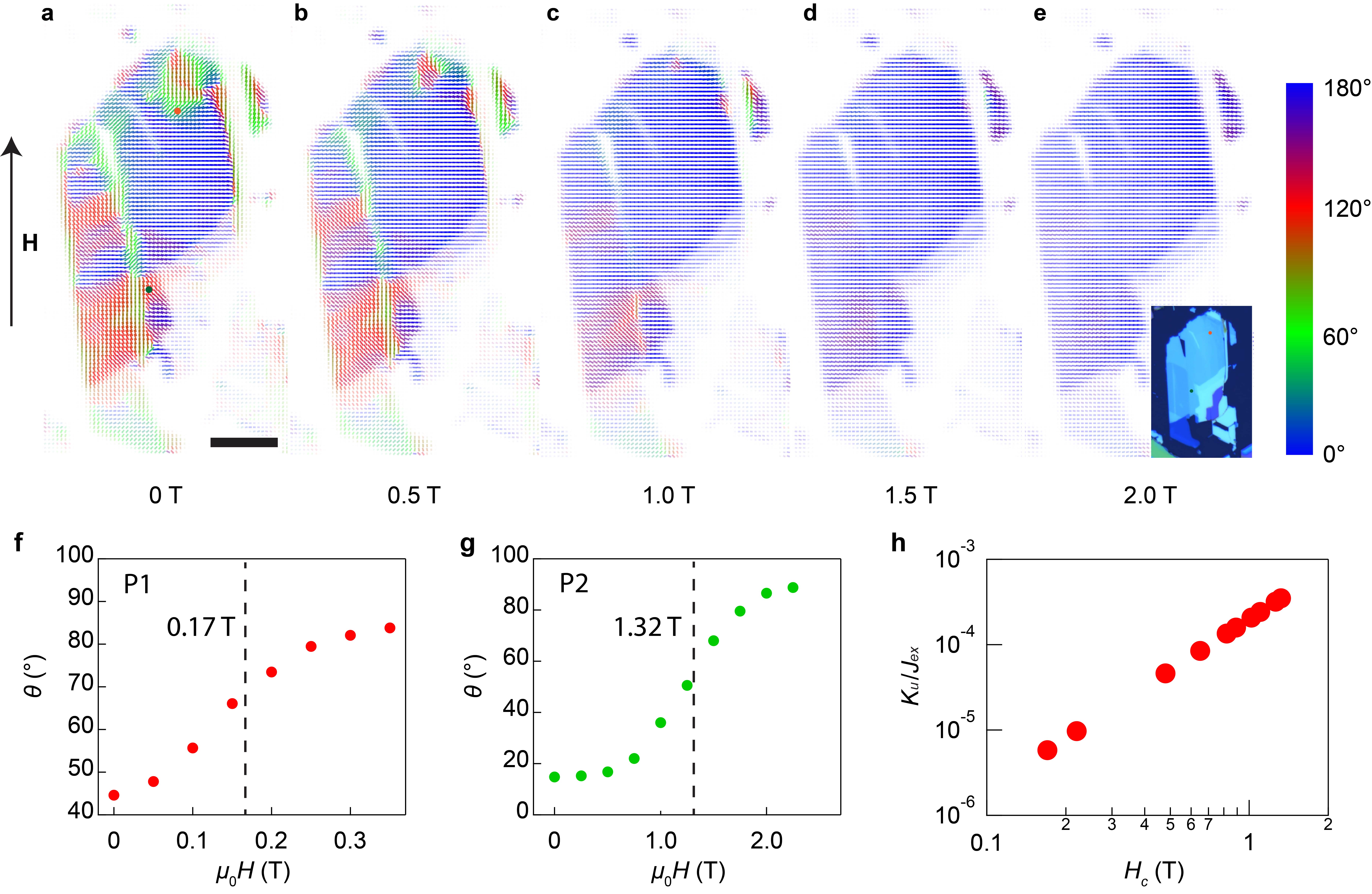}
\caption{\textbf{Mapping Néel Vectors under Different Magnetic Fields.} {\bf a--e}, Néel vector mapping of a MnPSe$_3$ thin flake (25 nm) under (a) 0 T, (b) 0.5 T, (c) 1.0 T, (d) 1.5 T, and (e) 2.0 T. Each point is represented by a segment, with color and orientation depicting the Néel vector orientation, while length and opacity indicate its magnitude. Scale bar: 10 $\mathrm{\mu m}$ {\bf f}, Magnetic field dependence of the Néel vector orientation $\theta$ at the point marked by the red dot in (a), with an estimated spin-flop transition at 0.17 T. {\bf g}, Magnetic field dependence of the Néel vector orientation $\theta$ at the point marked by the green dot in (a), with an estimated spin-flop transition at 1.32 T. {\bf h}, A summary of the critical magnetic field and calculated in-plane anisotropy energy for all 11 samples/points measured. Here J$_{ex}$ is from neutron scattering experiment, and K$_{u}$ is caculated by the spin-flop field and J$_{ex}$.}
\label{Fig3}
\end{figure*}

Next, we investigate the Néel vector distribution across the entire sample under the magnetic field. Since the in-plane anisotropy of the MnPSe$_3$ is highly affected by external strain, the net anisotropy distribution  at 0 T varies from point to point. Fig.\ref{Fig3}a displays the Néel vector distribution across a thin-flake sample at 0 T, measured by a  scanning SHG confocal microscope.  The Néel vector at each point is represented by a small segment, with the length and opacity of the segment indicating the magnitude of the Néel vector, and the direction and color representing the orientation of the Néel vector. A nonuniform distribution of Néel vector orientations is expected in MnPSe$_3$ under strain from the substrate\cite{ninatnano21}. We then measure the Néel vector distribution at different in-plane magnetic fields ranging from 0 T to 2.0 T, as illustrated in Fig.\ref{Fig3}b-e. At each point, the Néel vector gradually rotates towards the direction perpendicular to the magnetic field. However, the critical magnetic field (defined as the magnetic field that rotates the Néel vector halfway) varies at each point. Fig.\ref{Fig3}f and Fig.\ref{Fig3}g show the evolution of the Néel vector orientation as a function of the magnetic field at the red and green points marked in Fig.\ref{Fig3}a. The critical magnetic field at both points differs by an order of magnitude. The theoretical result of the critical magnetic field is $\sqrt{2(J_{ex}S^2+K_u)K_u}$, indicating that the zero field net anisotropy energy in the measured sample varies by at least two orders of magnitude. In Fig.\ref{Fig3}h, we plot the critical magnetic field and the calculated 0 T net anisotropy energy for 11 as-exfoliated samples and measured points.  The ratio between the anisotropy energy and the exchange energy $K_u/J_{ex}$ ranges from $4\times 10^{-6}$ to $4\times 10^{-4}$. The results demonstrate that the anisotropy of exfoliated MnPSe$_3$ is not intrinsic, but is highly sensitive to external strain. \\


\textbf{Magnetic field tunable Néel vector and temperature in the monolayer}

\begin{figure*}
\centering
\includegraphics[width=0.8\textwidth]{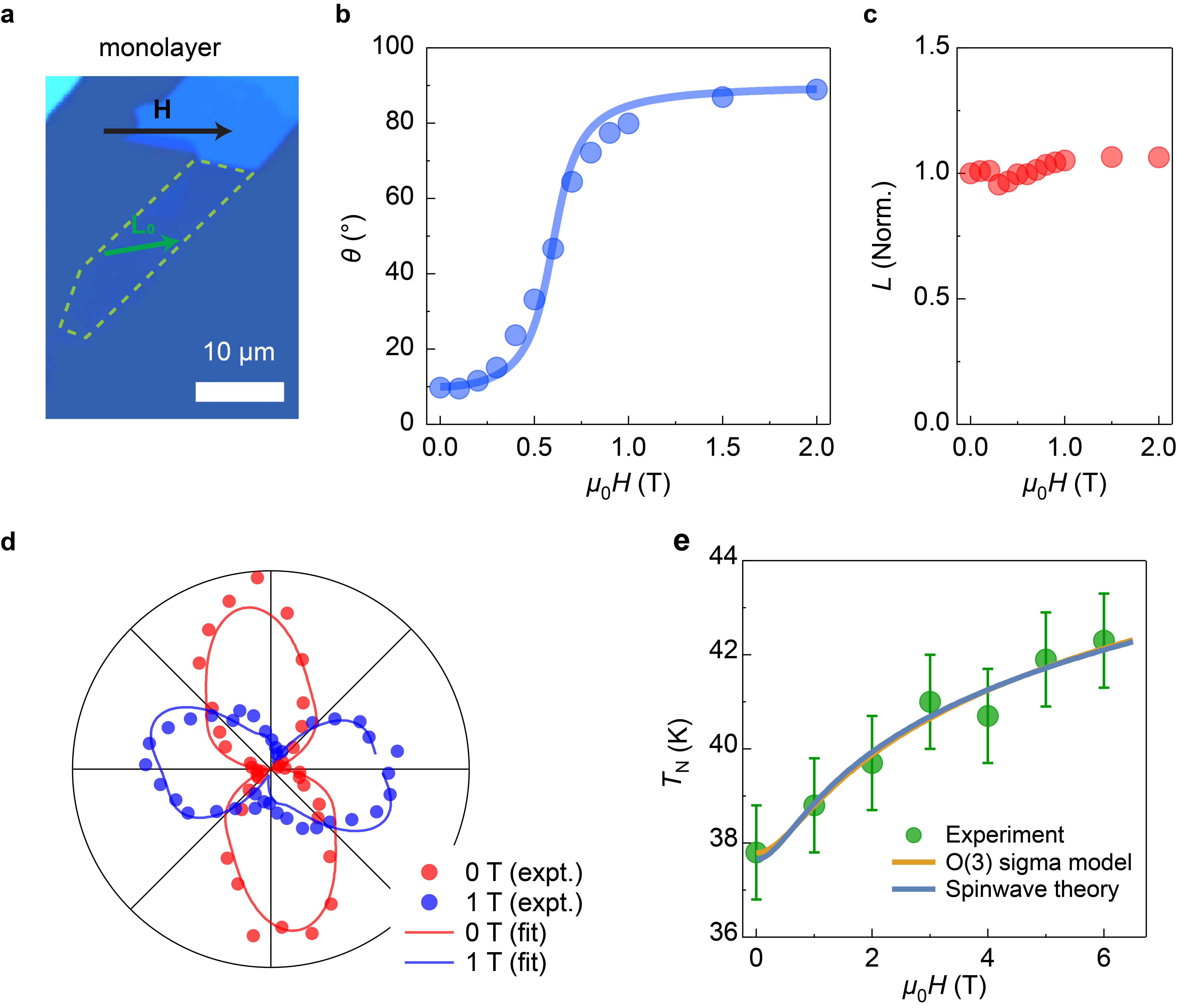}
\caption{\textbf{Magnetic Field Tuning of Monolayer MnPSe$_3$.} {\bf a}, Optical image of the monolayer sample (encircled by the dashed line). The Néel vector at zero field (green arrow) and the magnetic field direction (black arrow) are labeled. {\bf b}, Magnetic field dependence of the Néel vector orientation $\theta$. The blue circles represent experimental data, and the curve is derived from the simulation. {\bf c}, Magnetic field dependence of the Néel vector magnitude $L$. {\bf d}, \red{Polarization-dependent SHG polar patterns at different magnetic fields.} {\bf e}, Magnetic field dependence of the Néel temperature $T_N$. The green circles represent experimental data. The orange curve and the blue curve are fitted from the effective O(3) sigma model and spin-wave theory, respectively.}
\label{Fig4}
\end{figure*}

We proceed to examine the behavior of monolayer MnPSe$_3$ under a magnetic field. In a monolayer sample, the long-range ordering is considered more fragile than in thin-flake samples due to the absence of interlayer coupling. This is also evidenced by a significant drop in the Néel temperature from bilayer to monolayer samples\cite{ninatnano21}. Fig.\ref{Fig4}a shows the optical image of a monolayer sample (outlined by the green dashed line). We apply an inplane magnetic field, with the Néel vector at zero magnetic field situated around 7 degrees away from the magnetic field. Fig.\ref{Fig4}b presents the measured Néel vector orientation (circles) and the theoretical curve (blue line) as a function of the magnetic field.  Fig.\ref{Fig4}c depicts the extracted magnitude of the Néel vector as a function of the magnetic field, which is expected to remain constant. Another distinct behavior of the monolayer is that the magnetic field increases the Néel temperature. From 0 T to 6 T, there is a considerable increase of 5 K (12\%). In the thin-flake samples, the change is 0.3 K (see SI). We believe the presence of anisotropy from the magnetic field aids in stabilizing the long-range order and increasing the Néel temperature. For the monolayer without interlayer coupling, the long-range order is primarily stabilized by the weak external anisotropy. Therefore, the existence of anisotropy from the magnetic field can significantly increase the Néel temperature. For thicker samples, however, the long-range order is predominantly stabilized by interlayer coupling, so the magnetic field plays a less significant role.

\red{We now model the} field-induced enhancement of the Néel temperature. \red{The dominant energy scales of the system are $\mathrm{SO}(3)$-symmetric magnetic exchange interactions, with the intrinsic XY anisotropy estimated to be orders of magnitude smaller than exchange interactions \cite{calderprb2021,jana23}.
The} dynamics of the Néel vector \red{due to exchange interactions} can thus be described in terms of an effective $\mathrm{O}(3)$ sigma model.
However, magnetic ordering, i.e. the spontaneous breaking of the \emph{continuous} $\mathrm{O}(3)$ symmetry is forbidden at finite temperatures in $d=2$ spatial dimensions, with the correlation length $\xi$ becoming exponentially large at low temperatures.
\red{Instead, finite (effective) anisotropies $K_\mathrm{eff.}$ which break the $\mathrm{O}(3)$ symmetry are required for long-range order to occur at finite temperature. To obtain an estimate for the critical temperature, we argue that long-range order emerges} if the energetic gain due to the ordering, $\delta E$, is large compared to thermal fluctuations.
\red{Here,} the energy gain for a correlation area in two dimensions $\sim \xi^2$ can be estimated as $\delta E \sim \xi^2 K_\mathrm{eff.}$.
\red{Comparing with the energy scale $E_\mathrm{th} \sim k_\mathrm{B} T$ associated with thermal fluctuations, one finds that the magnetic phase transition occurs at a critical temperature $T$ which satisfies}
\begin{equation} \label{eq:dxiT}
	K_\mathrm{eff.} \xi^2 \sim  k_\mathrm{B} T.
\end{equation}
Here, $K_\mathrm{eff.}$ is some \emph{effective} anisotropy scale that is determined by a combination of the intrinsic anisotropies as well as the field-induced anisotropy as derived in Eq.~\eqref{eq_KH}.
The correlation length of the O(3) sigma model at finite temperatures $T$  can be written of the form (up to perturbative corrections) \cite{hasenfratz91,caracciolo95}
\begin{equation} \label{eq:xi-T}
	\xi \simeq C_\xi \mathrm{e}^{\frac{2\pi \rho}{k_\mathrm{B} T} }\,\frac{k_\mathrm{B} T}{2\pi \rho},
\end{equation}
where $C_\xi \approx 0.0124865$ is a numerical constant \cite{caracciolo95}, and $\rho = (J_1 - 6 J_2 + 4 J_3)S^2 /\sqrt{3}$ is the spin-wave stiffness of the  spin-$S$ Heisenberg model on the honeycomb lattice with $n$-th nearest neighbor interaction strengths $J_n$.
Note that Eq.~\eqref{eq:dxiT} is applicable if the correlation volume contains many spins.
Considering temperatures on the order of $T \simeq 35 \, \mathrm{K}$, we use the reported \cite{calderprb2021} exchange couplings $J_1 = 0.45 \, \mathrm{meV}$, $J_2 = 0.03 \, \mathrm{meV}$ and $J_3 = 0.19 \, \mathrm{meV}$ to find the correlation length $\xi \approx 3.72$. Hence, a correlation volume contains on the order of 10 spins, justifying our assumption.
Using Eq.~\eqref{eq:xi-T} in Eq.~\eqref{eq:dxiT} and expanding to leading order, the critical temperature as a function of the effective anisotropy is obtained as
\begin{equation} \label{eq:T-corrl}
	T \approx \frac{4 \pi \rho}{ k_\mathrm{B}} \frac{1}{C_0 - \ln \left( \frac{K_\mathrm{eff.}}{\rho} \right)},
\end{equation}
where $C_0$ is some undetermined constant. \red{As previously argued, the field-induced contribution to the anisotropy grows quadratically} with the applied field strength $K_\mathrm{eff} \sim K_0 + H^2$. \red{Thus}, an enhancement of the critical temperature $T_c$ becomes evident. Here, $K_0$ accounts for the undetermined effective anisotropic field from other sources, including the single-ion anisotropy and the strain effects.
\red{We note that the argument presented here relies on perturbing the near-critical O(3) sigma model with a single \emph{effective} anisotropy scale which combines intrinsic and field-induced contributions, while in a more microscopic treatment two independent anisotropy axes can be defined. We therefore complement} the above argument with a microscopic spin wave theory, where the critical $T$ is identified as the temperature where thermal order parameter fluctuations become large and disorder the system (see Methods section for details). The fitted results derived from both the O(3) sigma model and the microscopic spin-wave theory are depicted in Fig.\ref{Fig4}. Both models, yielding similar fitting curves, effectively capture the core elements underlying the increase in the Néel temperature (T$_N$) in response to an rise in the in-plane magnetic field.\\

\textbf{Discussion}

In conclusion, our study demonstrates the significant influence of an in-plane magnetic field as an effective anisotropy in governing spin orientations within the van der Waals easy-plane magnet MnPSe$_3$. We have successfully mapped the non-uniform but remarkably small in-plane anisotropy across the sample, further substantiating the crucial role of external anisotropy in determining spin order in this class of in-plane magnets. Notably, in monolayer systems, the field-induced anisotropy leads to an increase in the Néel temperature, which highlights the tunability of two-dimensional long-range order to external anisotropies.

\bibliography{references}

\begin{thebibliography}{10}
\expandafter\ifx\csname url\endcsname\relax
  \def\url#1{\texttt{#1}}\fi
\expandafter\ifx\csname urlprefix\endcsname\relax\def\urlprefix{URL }\fi
\providecommand{\bibinfo}[2]{#2}
\providecommand{\eprint}[2][]{\url{#2}}

\bibitem{baltzrmp2018}
\bibinfo{author}{Baltz, V.} \emph{et~al.}
\newblock \bibinfo{title}{{Antiferromagnetic spintronics}}.
\newblock \emph{\bibinfo{journal}{Rev. Mod. Phys.}}
  \textbf{\bibinfo{volume}{90}}, \bibinfo{pages}{015005}
  (\bibinfo{year}{2018}).
\newblock
  \urlprefix\url{https://link.aps.org/doi/10.1103/RevModPhys.90.015005}.

\bibitem{Palacioprb1980}
\bibinfo{author}{Palacio, F.}, \bibinfo{author}{Paduan-Filho, A.} \&
  \bibinfo{author}{Carlin, R.~L.}
\newblock \bibinfo{title}{{Phase diagram of antiferromagnetic
  ${\mathrm{K}}_{2}$[Fe${\mathrm{Cl}}_{5}$(${\mathrm{H}}_{2}$O)]}}.
\newblock \emph{\bibinfo{journal}{Phys. Rev. B}} \textbf{\bibinfo{volume}{21}},
  \bibinfo{pages}{296--298} (\bibinfo{year}{1980}).
\newblock \urlprefix\url{https://link.aps.org/doi/10.1103/PhysRevB.21.296}.

\bibitem{blazeypr1968}
\bibinfo{author}{Blazey, K.~W.} \& \bibinfo{author}{Rohrer, H.}
\newblock \bibinfo{title}{{Antiferromagnetism and the Magnetic Phase Diagram of
  GdAl${\mathrm{O}}_{3}$}}.
\newblock \emph{\bibinfo{journal}{Phys. Rev.}} \textbf{\bibinfo{volume}{173}},
  \bibinfo{pages}{574--580} (\bibinfo{year}{1968}).
\newblock \urlprefix\url{https://link.aps.org/doi/10.1103/PhysRev.173.574}.

\bibitem{jaccarinojmmm1983}
\bibinfo{author}{Jaccarino, V.}, \bibinfo{author}{King, A.},
  \bibinfo{author}{Motokawa, M.}, \bibinfo{author}{Sakakibara, T.} \&
  \bibinfo{author}{Date, M.}
\newblock \bibinfo{title}{{Temperature dependence of FeF$_2$ spin flop field}}.
\newblock \emph{\bibinfo{journal}{Journal of Magnetism and Magnetic Materials}}
  \textbf{\bibinfo{volume}{31-34}}, \bibinfo{pages}{1117--1118}
  (\bibinfo{year}{1983}).
\newblock
  \urlprefix\url{https://www.sciencedirect.com/science/article/pii/0304885383908223}.

\bibitem{jacobsjap1961}
\bibinfo{author}{Jacobs, I.~S.}
\newblock \bibinfo{title}{{Spin-Flopping in MnF$_2$ by High Magnetic Fields}}.
\newblock \emph{\bibinfo{journal}{Journal of Applied Physics}}
  \textbf{\bibinfo{volume}{32}}, \bibinfo{pages}{S61--S62}
  (\bibinfo{year}{1961}).
\newblock \urlprefix\url{https://doi.org/10.1063/1.2000500}.
\newblock \eprint{https://doi.org/10.1063/1.2000500}.

\bibitem{Jacobspr1967}
\bibinfo{author}{Jacobs, I.~S.} \& \bibinfo{author}{Lawrence, P.~E.}
\newblock \bibinfo{title}{{Metamagnetic Phase Transitions and Hysteresis in
  ${\mathrm{FeCl}}_{2}$}}.
\newblock \emph{\bibinfo{journal}{Phys. Rev.}} \textbf{\bibinfo{volume}{164}},
  \bibinfo{pages}{866--878} (\bibinfo{year}{1967}).
\newblock \urlprefix\url{https://link.aps.org/doi/10.1103/PhysRev.164.866}.

\bibitem{leenanoletter16}
\bibinfo{author}{Lee, J.-U.} \emph{et~al.}
\newblock \bibinfo{title}{{Ising-type magnetic ordering in atomically thin
  FePS$_3$}}.
\newblock \emph{\bibinfo{journal}{Nano Lett.}} \textbf{\bibinfo{volume}{16}},
  \bibinfo{pages}{7433--7438} (\bibinfo{year}{2016}).

\bibitem{ninatnano21}
\bibinfo{author}{Ni, Z.} \emph{et~al.}
\newblock \bibinfo{title}{{Imaging the N{\'e}el vector switching in the
  monolayer antiferromagnet {MnPSe$_3$} with strain-controlled Ising order}}.
\newblock \emph{\bibinfo{journal}{Nat. Nanotechnol.}}
  \textbf{\bibinfo{volume}{16}}, \bibinfo{pages}{782--787}
  (\bibinfo{year}{2021}).

\bibitem{Xzhangnanolett21}
\bibinfo{author}{Zhang, X.-X.} \emph{et~al.}
\newblock \bibinfo{title}{{Spin dynamics slowdown near the antiferromagnetic
  critical point in atomically thin FePS$_3$}}.
\newblock \emph{\bibinfo{journal}{Nano Lett.}} \textbf{\bibinfo{volume}{21}},
  \bibinfo{pages}{5045--5052} (\bibinfo{year}{2021}).

\bibitem{zhangnanolett21}
\bibinfo{author}{Zhang, Q.} \emph{et~al.}
\newblock \bibinfo{title}{{Observation of giant optical linear dichroism in a
  zigzag antiferromagnet FePS$_3$}}.
\newblock \emph{\bibinfo{journal}{Nano Lett.}} \textbf{\bibinfo{volume}{12}}
  (\bibinfo{year}{2021}).
\newblock \urlprefix\url{doi.org/10.1021/acs.nanolett.1c02188}.

\bibitem{songsci18}
\bibinfo{author}{Song, T.} \emph{et~al.}
\newblock \bibinfo{title}{{Giant tunneling magnetoresistance in spin-filter van
  der Waals heterostructures}}.
\newblock \emph{\bibinfo{journal}{Science}} \textbf{\bibinfo{volume}{360}},
  \bibinfo{pages}{1214--1218} (\bibinfo{year}{2018}).

\bibitem{kleinsci18}
\bibinfo{author}{Klein, D.~R.} \emph{et~al.}
\newblock \bibinfo{title}{{Probing magnetism in 2D van der Waals crystalline
  insulators via electron tunneling}}.
\newblock \emph{\bibinfo{journal}{Science}} \textbf{\bibinfo{volume}{360}},
  \bibinfo{pages}{1218--1222} (\bibinfo{year}{2018}).

\bibitem{wangnatmater21}
\bibinfo{author}{Wang, X.} \emph{et~al.}
\newblock \bibinfo{title}{{Spin-induced linear polarization of
  photoluminescence in antiferromagnetic van der Waals crystals}}.
\newblock \emph{\bibinfo{journal}{Nat. Mater.}}  (\bibinfo{year}{2021}).

\bibitem{longnanoletter20}
\bibinfo{author}{Long, G.} \emph{et~al.}
\newblock \bibinfo{title}{{Persistence of magnetism in atomically thin MnPS$_3$
  crystals}}.
\newblock \emph{\bibinfo{journal}{Nano Lett.}} \textbf{\bibinfo{volume}{20}},
  \bibinfo{pages}{2452--2459} (\bibinfo{year}{2020}).

\bibitem{basnetprm2021}
\bibinfo{author}{Basnet, R.}, \bibinfo{author}{Wegner, A.},
  \bibinfo{author}{Pandey, K.}, \bibinfo{author}{Storment, S.} \&
  \bibinfo{author}{Hu, J.}
\newblock \bibinfo{title}{Highly sensitive spin-flop transition in
  antiferromagnetic van der waals material $m\mathrm{P}{\mathrm{s}}_{3}$
  ($m=\mathrm{Ni}$ and mn)}.
\newblock \emph{\bibinfo{journal}{Phys. Rev. Mater.}}
  \textbf{\bibinfo{volume}{5}}, \bibinfo{pages}{064413} (\bibinfo{year}{2021}).
\newblock
  \urlprefix\url{https://link.aps.org/doi/10.1103/PhysRevMaterials.5.064413}.

\bibitem{feringaprb2022}
\bibinfo{author}{Feringa, F.}, \bibinfo{author}{Vink, J.~M.} \&
  \bibinfo{author}{van Wees, B.~J.}
\newblock \bibinfo{title}{Spin-flop transition in the quasi-two-dimensional
  antiferromagnet ${\mathrm{mnps}}_{3}$ detected via thermally generated magnon
  transport}.
\newblock \emph{\bibinfo{journal}{Phys. Rev. B}}
  \textbf{\bibinfo{volume}{106}}, \bibinfo{pages}{224409}
  (\bibinfo{year}{2022}).
\newblock \urlprefix\url{https://link.aps.org/doi/10.1103/PhysRevB.106.224409}.

\bibitem{mcguireprm2017}
\bibinfo{author}{McGuire, M.~A.} \emph{et~al.}
\newblock \bibinfo{title}{{Magnetic behavior and spin-lattice coupling in
  cleavable van der Waals layered ${\mathrm{CrCl}}_{3}$ crystals}}.
\newblock \emph{\bibinfo{journal}{Phys. Rev. Mater.}}
  \textbf{\bibinfo{volume}{1}}, \bibinfo{pages}{014001} (\bibinfo{year}{2017}).
\newblock
  \urlprefix\url{https://link.aps.org/doi/10.1103/PhysRevMaterials.1.014001}.

\bibitem{cainanoletter19}
\bibinfo{author}{Cai, X.} \emph{et~al.}
\newblock \bibinfo{title}{{Atomically thin CrCl$_3$: an in-plane layered
  antiferromagnetic insulator}}.
\newblock \emph{\bibinfo{journal}{Nano Lett.}} \textbf{\bibinfo{volume}{19}},
  \bibinfo{pages}{3993--3998} (\bibinfo{year}{2019}).

\bibitem{macneillprl2019}
\bibinfo{author}{MacNeill, D.} \emph{et~al.}
\newblock \bibinfo{title}{{Gigahertz Frequency Antiferromagnetic Resonance and
  Strong Magnon-Magnon Coupling in the Layered Crystal ${\mathrm{CrCl}}_{3}$}}.
\newblock \emph{\bibinfo{journal}{Phys. Rev. Lett.}}
  \textbf{\bibinfo{volume}{123}}, \bibinfo{pages}{047204}
  (\bibinfo{year}{2019}).
\newblock
  \urlprefix\url{https://link.aps.org/doi/10.1103/PhysRevLett.123.047204}.

\bibitem{wangnatnano2019}
\bibinfo{author}{Wang, Z.} \emph{et~al.}
\newblock \bibinfo{title}{{Determining the phase diagram of atomically thin
  layered antiferromagnet CrCl$_3$}}.
\newblock \emph{\bibinfo{journal}{Nature Nanotechnology}}
  \textbf{\bibinfo{volume}{14}}, \bibinfo{pages}{1116--1122}
  (\bibinfo{year}{2019}).

\bibitem{amilcarscience2021}
\bibinfo{author}{Bedoya-Pinto, A.} \emph{et~al.}
\newblock \bibinfo{title}{{Intrinsic 2D-XY ferromagnetism in a van der Waals
  monolayer}}.
\newblock \emph{\bibinfo{journal}{Science}} \textbf{\bibinfo{volume}{374}},
  \bibinfo{pages}{616--620} (\bibinfo{year}{2021}).
\newblock
  \urlprefix\url{https://www.science.org/doi/abs/10.1126/science.abd5146}.
\newblock \eprint{https://www.science.org/doi/pdf/10.1126/science.abd5146}.

\bibitem{kleinnatphys19}
\bibinfo{author}{Klein, D.~R.} \emph{et~al.}
\newblock \bibinfo{title}{{Enhancement of interlayer exchange in an ultrathin
  two-dimensional magnet}}.
\newblock \emph{\bibinfo{journal}{Nat. Phys.}} \textbf{\bibinfo{volume}{15}},
  \bibinfo{pages}{1255--1260} (\bibinfo{year}{2019}).
\newblock \urlprefix\url{https://doi.org/10.1038/s41567-019-0651-0}.

\bibitem{wiedenmannssc81}
\bibinfo{author}{Wiedenmann, A.}, \bibinfo{author}{Rossat-Mignod, J.},
  \bibinfo{author}{Louisy, A.}, \bibinfo{author}{Brec, R.} \&
  \bibinfo{author}{Rouxel, J.}
\newblock \bibinfo{title}{{Neutron diffraction study of the layered compounds
  MnPSe$_3$ and FePSe$_3$}}.
\newblock \emph{\bibinfo{journal}{Solid State Commun.}}
  \textbf{\bibinfo{volume}{40}}, \bibinfo{pages}{1067--1072}
  (\bibinfo{year}{1981}).

\bibitem{jeevanandamjpcm1999}
\bibinfo{author}{Jeevanandam, P.} \& \bibinfo{author}{Vasudevan, S.}
\newblock \bibinfo{title}{{Magnetism in MnPSe$_3$: a layered 3d5
  antiferromagnet with unusually large XY anisotropy}}.
\newblock \emph{\bibinfo{journal}{Journal of Physics: Condensed Matter}}
  \textbf{\bibinfo{volume}{11}}, \bibinfo{pages}{3563} (\bibinfo{year}{1999}).
\newblock \urlprefix\url{https://dx.doi.org/10.1088/0953-8984/11/17/314}.

\bibitem{wildesprb2006}
\bibinfo{author}{Wildes, A.}, \bibinfo{author}{R{\o}nnow, H.},
  \bibinfo{author}{Roessli, B.}, \bibinfo{author}{Harris, M.} \&
  \bibinfo{author}{Godfrey, K.}
\newblock \bibinfo{title}{{Static and dynamic critical properties of the
  quasi-two-dimensional antiferromagnet MnPS$_3$}}.
\newblock \emph{\bibinfo{journal}{Phys. Rev. B}} \textbf{\bibinfo{volume}{74}},
  \bibinfo{pages}{094422} (\bibinfo{year}{2006}).

\bibitem{ressoucheprb10}
\bibinfo{author}{Ressouche, E.} \emph{et~al.}
\newblock \bibinfo{title}{{Magnetoelectric ${\text{MnPS}}_{3}$ as a candidate
  for ferrotoroidicity}}.
\newblock \emph{\bibinfo{journal}{Phys. Rev. B}} \textbf{\bibinfo{volume}{82}},
  \bibinfo{pages}{100408} (\bibinfo{year}{2010}).

\bibitem{lanconprb16}
\bibinfo{author}{Lancon, D.} \emph{et~al.}
\newblock \bibinfo{title}{{Magnetic structure and magnon dynamics of the
  quasi-two-dimensional antiferromagnet ${\mathrm{FePS}}_{3}$}}.
\newblock \emph{\bibinfo{journal}{Phys. Rev. B}} \textbf{\bibinfo{volume}{94}},
  \bibinfo{pages}{214407} (\bibinfo{year}{2016}).
\newblock \urlprefix\url{https://link.aps.org/doi/10.1103/PhysRevB.94.214407}.

\bibitem{Maisciadv2021}
\bibinfo{author}{Mai, T.~T.} \emph{et~al.}
\newblock \bibinfo{title}{{Magnon-phonon hybridization in 2D antiferromagnet
  MnPSe$_3$}}.
\newblock \emph{\bibinfo{journal}{Science Advances}}
  \textbf{\bibinfo{volume}{7}}, \bibinfo{pages}{eabj3106}
  (\bibinfo{year}{2021}).
\newblock
  \urlprefix\url{https://www.science.org/doi/abs/10.1126/sciadv.abj3106}.
\newblock \eprint{https://www.science.org/doi/pdf/10.1126/sciadv.abj3106}.

\bibitem{calderprb2021}
\bibinfo{author}{Calder, S.}, \bibinfo{author}{Haglund, A.~V.},
  \bibinfo{author}{Kolesnikov, A.~I.} \& \bibinfo{author}{Mandrus, D.}
\newblock \bibinfo{title}{{Magnetic exchange interactions in the van der Waals
  layered antiferromagnet $\mathrm{Mn}\mathrm{P}{\mathrm{Se}}_{3}$}}.
\newblock \emph{\bibinfo{journal}{Phys. Rev. B}}
  \textbf{\bibinfo{volume}{103}}, \bibinfo{pages}{024414}
  (\bibinfo{year}{2021}).
\newblock \urlprefix\url{https://link.aps.org/doi/10.1103/PhysRevB.103.024414}.

\bibitem{YanAM2023}
\bibinfo{author}{Liu, C.} \emph{et~al.}
\newblock \bibinfo{title}{{Probing the Néel-Type Antiferromagnetic Order and
  Coherent Magnon–Exciton Coupling in Van Der Waals VPS$_3$}}.
\newblock \emph{\bibinfo{journal}{Advanced Materials}}
  \textbf{\bibinfo{volume}{35}}, \bibinfo{pages}{2300247}
  (\bibinfo{year}{2023}).
\newblock
  \urlprefix\url{https://onlinelibrary.wiley.com/doi/abs/10.1002/adma.202300247}.
\newblock
  \eprint{https://onlinelibrary.wiley.com/doi/pdf/10.1002/adma.202300247}.

\bibitem{basnet22}
\bibinfo{author}{Basnet, R.} \emph{et~al.}
\newblock \bibinfo{title}{{Controlling magnetic exchange and anisotropy by
  nonmagnetic ligand substitution in layered $M\mathrm{P}{X}_{3}$
  ($M=\mathrm{Ni}$, Mn; $X=\mathrm{S}$, Se)}}.
\newblock \emph{\bibinfo{journal}{Phys. Rev. Res.}}
  \textbf{\bibinfo{volume}{4}}, \bibinfo{pages}{023256} (\bibinfo{year}{2022}).
\newblock
  \urlprefix\url{https://link.aps.org/doi/10.1103/PhysRevResearch.4.023256}.

\bibitem{jana23}
\bibinfo{author}{Jana, D.} \emph{et~al.}
\newblock \bibinfo{title}{{Magnon gap excitations in van der Waals
  antiferromagnet MnPSe$_3$}} (\bibinfo{year}{2023}).
\newblock \eprint{2309.06866}.

\bibitem{fiebigprl94}
\bibinfo{author}{Fiebig, M.}, \bibinfo{author}{Fr{\"o}hlich, D.},
  \bibinfo{author}{Krichevtsov, B.} \& \bibinfo{author}{Pisarev, R.~V.}
\newblock \bibinfo{title}{{Second harmonic generation and
  magnetic-dipole-electric-dipole interference in antiferromagnetic
  {Cr$_2$O$_3$}}}.
\newblock \emph{\bibinfo{journal}{Phys. Rev. Lett.}}
  \textbf{\bibinfo{volume}{73}}, \bibinfo{pages}{2127} (\bibinfo{year}{1994}).

\bibitem{niprl2021}
\bibinfo{author}{Ni, Z.} \emph{et~al.}
\newblock \bibinfo{title}{{Direct Imaging of Antiferromagnetic Domains and
  Anomalous Layer-Dependent Mirror Symmetry Breaking in Atomically Thin
  ${\mathrm{MnPS}}_{3}$}}.
\newblock \emph{\bibinfo{journal}{Phys. Rev. Lett.}}
  \textbf{\bibinfo{volume}{127}}, \bibinfo{pages}{187201}
  (\bibinfo{year}{2021}).
\newblock
  \urlprefix\url{https://link.aps.org/doi/10.1103/PhysRevLett.127.187201}.

\bibitem{chuprl20}
\bibinfo{author}{Chu, H.} \emph{et~al.}
\newblock \bibinfo{title}{{Linear magnetoelectric phase in ultrathin
  ${\mathrm{MnPS}}_{3}$ probed by optical second harmonic generation}}.
\newblock \emph{\bibinfo{journal}{Phys. Rev. Lett.}}
  \textbf{\bibinfo{volume}{124}}, \bibinfo{pages}{027601}
  (\bibinfo{year}{2020}).

\bibitem{hasenfratz91}
\bibinfo{author}{Hasenfratz, P.} \& \bibinfo{author}{Niedermayer, F.}
\newblock \bibinfo{title}{{The exact correlation length of the
  antiferromagnetic d=2+1 Heisenberg model at low temperatures}}.
\newblock \emph{\bibinfo{journal}{Physics Letters B}}
  \textbf{\bibinfo{volume}{268}}, \bibinfo{pages}{231--235}
  (\bibinfo{year}{1991}).
\newblock
  \urlprefix\url{https://www.sciencedirect.com/science/article/pii/0370269391908095}.

\bibitem{caracciolo95}
\bibinfo{author}{Caracciolo, S.}, \bibinfo{author}{Edwards, R.~G.},
  \bibinfo{author}{Pelissetto, A.} \& \bibinfo{author}{Sokal, A.~D.}
\newblock \bibinfo{title}{{Asymptotic Scaling in the Two-Dimensional O(3)
  \ensuremath{\sigma} Model at Correlation Length 1${0}^{5}$}}.
\newblock \emph{\bibinfo{journal}{Phys. Rev. Lett.}}
  \textbf{\bibinfo{volume}{75}}, \bibinfo{pages}{1891--1894}
  (\bibinfo{year}{1995}).
\newblock \urlprefix\url{https://link.aps.org/doi/10.1103/PhysRevLett.75.1891}.

\end{thebibliography}

\

\section{Methods}

\subsection*{Sample Preparation}
 Single crystals of MnPSe$_3$ were grown by the chemical vapor transport method. Elemental powders of high purity Mn, P, and Se were pressed into a pellet and sealed inside a quartz tube under vacuum. The tube was then annealed for a week at 730 $^\circ$C to form polycrystalline MnPSe$_3$ power, the composition of which was verified with powder X-ray diffraction. Crystals were then grown using the chemical vapor transport method with iodine as transport agent: 2 g of the powder and 0.4 g of iodine crystals were placed at the end of a quartz tube, which was sealed off at 13 cm length under vacuum. The sealed tube was then set in a temperature gradient of 650/525 $^\circ$C for four days to transport the starting materials placed at the hot end to the cold end. The Mn : P : Se ratio was measured to be 1.00(1) : 0.96(1) : 3.07(1) with energy-dispersive X-ray spectroscopy. The ultra-thin samples were prepared by a standard mechanical exfoliation process  on Si substrates with 90 nm thick SiO$_2$ from a few MnPSe$_3$ bulk crystals with $T_N$ $\sim$ 68 K. The samples were put into a vacuum environment after exfoliation. The total exposure time in air for samples exfoliated in a glove box is less than one minute before loading into the cryostat where the sample is under vacuum.

 \subsection*{\red{Scanning} SHG microscopy}
\red{All the experimental data are measured under the same experimental settings.} The sample was situated on a metal platform within a closed-cycle cryostat, with the platform's temperature managed by a localized heater and the in-plane magnetic field added by a superconducting magnet. An ultrafast 800-nm Ti-sapphire laser pulse, with an approximate duration of 50 fs at a repetition rate of 80 MHz, was focused onto a 2 $\mu$m beam spot on the sample at normal incidence by a 50X objective. A typical laser power of 200 $\mu$W was utilized \red{for thick samples and 100 $\mu$W for monolayer samples. No sample degradation was observed after the measurements.} The reflected SHG light was captured by the same objective, then reflected by either a dichroic mirror into a photomultiplier tube. A photon counter synchronized to the 80 MHz laser frequency was used to isolate the signal counts from the noise. The sensitivity of the photo counters is 0.2 counts per second (c.p.s.). The polarization of the fundamental light was controlled by a half-wave plate, while the polarization of the second-harmonic light was analyzed using another linear polarizer.

\red{SHG imaging was executed by scanning the sample with three Attocube nanopositioners. For the microscopy imaging in Fig.\ref{Fig3}, SHG intensity image in the crossed configuration at ten different polarization states (0 to 180 degrees with a step of 20 degrees) were first measured. Then the ten intensity images were combined to fit the N\'eel vector orientation at each positions. For all other single-point SHG measurement, the positions were fixed while the SHG polar patterns were measured at different polarization states.}

 \subsection*{Spin-wave theory for the O(3) sigma model with anisotropies}

\red{We here present a spin-wave theory that demonstrates the enhancement of the Néel temperature as a function of the applied magnetic field. This theory incorporates two distinct anisotropy axis and is complementary to the argument given in the main text.}

\red{First, we note} that the microscopic Hamiltonian can be split into exchange and (local) anisotropy contributions, $\mathcal{H} = \mathcal{H}_\mathrm{ex.} + \mathcal{H}_\mathrm{aniso.}$. \red{In terms of the Néel vector $\vec n$}, the latter can be written of the form 
\begin{multline} \label{eq:h-aniso}
	\mathcal{H}_\mathrm{aniso.} = S^2 \sum_{i \in \mathrm{u.c.}} \big[-2 K_u (n^x \cos \beta + n^y \sin \beta)^2 \\ -2 K_H (n^y)^2 + 2K_z (n^z)^2 \big],
\end{multline}
where $K_z > 0$ is an easy-plane anisotropy, and $K_u>0$ is an easy-axis anisotropy which fixes the Néel-vector direction along an in-plane axis with angle $\beta$, and $K_H$ is the field-induced anisotropy (here, w.l.o.g., along the $\hat{y}$-axis).
\red{The anisotropy term can be written as a bilinear term} $\mathcal{H}_\mathrm{aniso.} = 2 S^2 \sum_i \vec{n}_i^\top \cdot K \cdot \vec{n}_i$, with the matrix
\red{\begin{equation}
    K = \begin{pmatrix}
        -2 K_u \cos^2 \beta & - K_u \sin 2 \beta & 0 \\
        - K_u \sin 2 \beta & -2 K_H \sin^2 \beta - 2 K_h & 0\\
        0 & 0 & 2 K_z
    \end{pmatrix}.
\end{equation}
}
\red{Upon diagonalizing $K$, the eigenvectors determine the principal anisotropy axes. Transforming the Néel vector into the eigenframe of $K$ (we denote the transformed fields as $\tilde{n}^x$ and $\tilde{n}^z$), the anisotropy Hamiltonian can be written as}
\begin{equation}
	\tilde{\mathcal{H}}_\mathrm{aniso.} = 2S^2 \sum_{i \in \mathrm{u.c.}} \left[ K_1 (\tilde{n}^x)^2 + K_2 (\tilde{n}^z)^2 \right] + \mathrm{const.},  
\end{equation}
where $K_1 = \sqrt{K_u^2 + K_H^2 - 2 K_u K_H \cos 2 \beta}>0$ and $K_2 = K_z + \frac{1}{2} \left( K_x + K_h + \sqrt{K_u^2 + K_H^2 - 2 K_u K_H \cos 2 \beta} \right) > 0$ are the anisotropy energy scales along the two principal axes. \red{Here, we have exploited the normalization $\vec{n}\cdot \vec{n} = 1$ which implies that only two independent anisotropies exist and allows us to subtract an overall constant to ensure that $K_1 > 0$ and $K_2>0$}.
Note that the $\mathrm{SO}(3)$-symmetric exchange interactions $\mathcal{H}_\mathrm{ex.}$ are invariant under the transformation to the eigenframe.
Hence, in this frame, $\tilde{\mathcal{H}}_\mathrm{aniso.}$ implies that the Néel vector will order along the $\tilde{y}$-axis.
\red{To investigate the stability of the ordered state against thermal fluctuations, we now perform a spin-wave analysis. To this end, we} parametrize order parameter fluctuations as 
\begin{equation}
	\vec{\tilde{n}}  = \begin{pmatrix}
		\tilde{n}^x  \\\sqrt{1-(\tilde{n}^x)^2 - (\tilde{n}^z)^2} \\  \tilde{n}^z 
	\end{pmatrix} \approx
	\begin{pmatrix}
		\tilde{n}^x \\ 1- \frac{1}{2} \left( (\tilde{n}^x)^2  +  (\tilde{n}^z)^2 \right) \\  \tilde{n}^z
	\end{pmatrix}
\end{equation}
In the ordered state, and after coarse graining, the $\mathrm{SO}(3)$-symmetric exchange interactions give rise to a stiffness term for the transverse fluctuations,
\begin{equation}
	\tilde{H}_\rho = \frac{\rho}{2} \int \mathrm{d}^2 \bvec{x} \left[ (\nabla \tilde{n}^x)^2 + (\nabla \tilde{n}^z)^2 \right],
\end{equation}
with the anisotropies giving rise mass terms for the $\tilde{n}^x$ and $\tilde{n}^z$ terms,
\begin{equation}
	\mathcal{H}_\mathrm{aniso.} = \frac{1}{2} \int \mathrm{d}^2 \bvec{x} \left[ \frac{8 S^2 K_1}{\sqrt{3}} (\tilde{n}^x)^2 + \frac{8 S^2 K_2}{\sqrt{3}} (\tilde{n}^z)^2 \right].
\end{equation}
\red{After Fourier-transforming, we can read off the correlation functions as}
$\langle n^\mu(\bvec{x}) n^\nu(\bvec{y}) \rangle = \delta_{\mu,\nu} (k_\mathrm{B} T) \int \frac{\mathrm{d}^2 \bvec{k}}{(2\pi)^2} \frac{\mathrm{e}^{\mathrm{i} \bvec{k} \cdot (\bvec{x} - \bvec{y})}}{\rho \bvec{k}^2 + m_\mu^2}$ where $m_\mu^2 = 8 S^2 K_\mu / \sqrt{3}$ for $\mu=x(1),z(2)$.
\red{From this, the magnitude of the transverse fluctuations is given by}
\begin{equation} \label{eq:nxny-corrls}
	\langle \left(\tilde{n}^x(\bvec{x}) \right)^2 + \left(\tilde{n}^z(\bvec{x}) \right)^2   \rangle \approx  \frac{k_\mathrm{B} T}{4 \pi \rho} \left[\ln \left(\frac{\rho}{K_1} \right) + \ln\left(\frac{\rho}{K_2} \right) + C_0  \right],
\end{equation}
where $C$ is some constant that involves an UV cutoff that is used to regularize the momentum-integration, and we have dropped terms $\rho/m^2 \ll 1$ in the logarithms.
The antiferromagnetic order is destabilized when the magnitude of transverse fluctuations \red{become} of order one, $\langle (\tilde{n}^x)^2 + (\tilde{n}^z)^2 \rangle \approx C_1 = \mathcal{O}(1)$. Using Eq.~\eqref{eq:nxny-corrls}, we then obtain the critical temperature as
\begin{equation} \label{eq:T-sw}
	T \approx \frac{4\pi \rho }{k_\mathrm{B}} \frac{C_1}{C_0 - \ln \frac{K_1 K_2}{\rho^2}},
\end{equation}
where $C_0$ is an undetermined constant near 1. The above equation is seen to correspond to a generalization of Eq.~\eqref{eq:T-corrl}. In this model, $K_z$, $K_u$ and $\rho$ are known. Only $\tilde{C}_0$ and $x$ are fitting parameters.

\section{Acknowledgments}
L.W. acknowledges support by the US Office of Naval Research (ONR) through the grant N00014-24-1-2064.  The development of the scanning imaging microscope was sponsored by the Army Research Office and was accomplished under Grant Number W911NF-20-2-0166 and W911NF-21-1-0131, and the University Research Foundation.  The sample exfoliation setup is based upon work supported by the Air Force Office of Scientific Research under award number FA9550-22-1-0449. Z.N. acknowledges support from Vagelos Institute of Energy  Science  and  Technology  graduate  Fellowship and Dissertation Completion Fellowship  at  the  University  of  Pennsylvania. D.G.M acknowledges support from the Gordon and Betty Moore Foundation's EPiQS Initiative, Grant GBMF9069.
\red{U.F.P.S. was supported by the Deutsche Forschungsgemeinschaft (DFG, German Research Foundation) through a Walter Benjamin fellowship, Project ID No. 449890867 and the DOE office of BES, through award number DE-SC0020305. L.B. was supported by the NSF CMMT program under Grant No. DMR-2116515, the Gordon and Betty Moore Foundation through Grant GBMF8690, and by the Simons Collaboration on Ultra-Quantum Matter, which is a grant from the Simons Foundation (651440).}
This research was supported in part by grant NSF PHY-1748958 to the Kavli Institute for Theoretical Physics (KITP).


\section{Addendum}

\textit{Data availability:} All data needed to evaluate the conclusions in the paper are present in the paper and the Supplementary Information. Additional data related to this paper could be requested from the authors.

\textit{Competing Interests: }The authors declare that they have no
competing financial interests.

\textit{Correspondence: }Correspondence and requests for materials
should be addressed to L.W. (liangwu@sas.upenn.edu)

\newpage

\end{document}